\documentclass[jgrga]{AGUTeX}
\usepackage{graphicx}
\usepackage{graphics}
\usepackage{amssymb,amsmath}
\usepackage{chngcntr}

\pdfpageattr {/Group << /S /Transparency /I true /CS /DeviceRGB>>}

\newcommand{\degree}{\ensuremath{^\circ}}

\authorrunninghead{WILSON et al.}
\titlerunninghead{}

\begin{document}

%
%
\title{{
Evidence for explosive silicic volcanism on the Moon from the extended distribution of thorium near the Compton-Belkovich Volcanic Complex
}
%
%
%
%
}

%
%
\authors{
  J.T. Wilson \altaffilmark{1}, 
V.R. Eke \altaffilmark{1}, R.J. Massey\altaffilmark{1},
R.C. Elphic\altaffilmark{2}, B.L. Jolliff\altaffilmark{3}, D.J. Lawrence\altaffilmark{4}, E.W. Llewellin\altaffilmark{5}, J.N. McElwaine\altaffilmark{5},
L.F.A. Teodoro\altaffilmark{6}}
\altaffiltext{1}{Institute for Computational Cosmology, Department of Physics, Durham University, Science Laboratories, South Road, Durham DH1 3LE, UK}
\altaffiltext{2}{Planetary Systems Branch,NASA Ames Research Center,
  MS 2453, Moffett Field, CA,94035-1000, USA}
\altaffiltext{3}{Department of Earth and Planetary Sciences and the McDonnell Centre for the Space Sciences, Washington University, One Brookings Drive, St Louis, Missouri 63130, USA}
\altaffiltext{4}{The Johns Hopkins University Applied Physics Laboratory, Laurel, Maryland, USA}
\altaffiltext{5}{Department of Earth Sciences, Durham University, Science Laboratories, South Road, Durham DH1 3LE, UK}
\altaffiltext{6}{BAER, Planetary Systems Branch, Space Sciences and Astrobiology Division, MS 245-3, NASA Ames Research Center, Moffett Field, CA 94035-1000, USA}

\begin{article}

%
%
\abstract{
%
%
We reconstruct the abundance of thorium near the Compton-Belkovich Volcanic Complex on the Moon, using data from the Lunar Prospector Gamma Ray Spectrometer.
We enhance the resolution via a pixon image reconstruction technique, and find that the thorium is distributed over a larger ($40\,\mathrm{km}\times 75$\,km) area than the ($25\,\mathrm{km}\times 35$\,km) high albedo region normally associated with Compton-Belkovich.
Our reconstructions show that inside this region, the thorium concentration is $14\!-\!26$\,ppm.
We also find additional thorium, spread up to $300$\,km eastward of the complex at $\sim\!2$\,ppm. 
The thorium must have been deposited during the formation of the volcanic complex, because subsequent lateral transport mechanisms, such as small impacts, are unable to move sufficient material. The morphology of the feature is consistent with pyroclastic dispersal and we conclude that the present distribution of thorium was likely created by the explosive eruption of silicic magma.}
\vspace{1cm}
%
%

\section{Introduction}\label{sec:Int}
\subsection{Gamma ray spectroscopy}
The chemical composition of the Moon's surface was
mapped by the Lunar Prospector spacecraft using gamma ray and neutron
spectroscopy \citep{Elphic98,Elphic2000,Feldman98,Feldman2000,Feldman2002,Lawrence1998,Lawrence2000,Lawrence2002,Prettyman2006} and these maps have led to an improved
understanding of the formation and evolution of the lunar surface and
interior \citep{Jolliff2000,Hagerty2006,Hagerty2009}. Both of these methods of measuring
elemental composition have the advantage, over other forms of
spectroscopy, of not being sensitive to the mineral form in which the
elements occur and of being able to probe composition at depths of a few 
tens of cm rather than only the top few wavelengths, as in ultraviolet to near-infrared reflectance spectroscopy. 
Further, in the case of gamma ray detection from the
natural decay of thorium (Th), uranium and potassium, the
inferred abundances do not depend on cosmic ray flux or ground truth
but only on having an accurate background subtraction
\citep{Metzger77} (though in practice bias may be introduced if the contribution of major elements to the background is not taken into account, see \citet{Prettyman2006} for details).  These chemical elements are particularly
interesting as they have large ionic radii and are incompatible so preferentially partition into the melt phase during magmagenesis, and remain in the
melt phase as it crystallizes. Thus the distribution of these elements acts as a
tracer of magmatic activity and differentiation.

Of the three chemical elements detectable from orbit, Th is the most
easily observed because its $2.61$\,MeV peak in the Moon's gamma ray spectrum
is both strong and well separated from other peaks \citep{Reedy78}.
Examination of Th abundance maps, along with other data, gave rise to the 
%
%
interpretation that 
the lunar surface comprises three terranes 
\citep{Jolliff2000}:  the low-Th Feldspathic Highland Terrane; the moderate-Th South Pole-Aitken basin; and
the high-Th Procellarum KREEP Terrane 
(named
after the materials with high potassium (K), rare earth element (REE) and
phosphorus (P) abundances that cover much of its surface but which also
contain other incompatible elements including Th \citep{WW79}).

%
%
%
Several anomalous regions of the Moon's surface fall outside this broad classification scheme, most notably a small but distinct Th enrichment located between the craters Compton (103.8°\degree E, 55.3\degree N) and Belkovich (90.2\degree E, 61.1\degree N) on the lunar farside. An isolated enrichment of Th was first detected at (60\degree N, 100\degree E) in the Lunar Prospector Gamma Ray Spectrometer (LP-GRS) data \citep{Lawrence99,Lawrence2000,Gillis2002}. 
\citet{Jolliff2011} associated this compositionally unique feature with a $\sim\!25\,\mathrm{km}\times 35$\,km topographically elevated, high albedo region that contains irregular depressions, cones and domes of varying size. They interpreted this region as a small, silicic volcanic complex, which they referred to as the Compton-Belkovich
Volcanic Complex (CBVC) \citep{Jolliff2011}. Uniquely, the high Th region of the CBVC is not coincident with an elevated FeO terrain (as would be expected from a KREEP basalt); instead the CBVC appears to have a low FeO abundance ($\sim 4-5\,\mathrm{wt.}\%$) that is similar to much of the lunar highlands \citep{Lawrence99,Jolliff2011}.
%
%
Crater counting results indicate a 
%
%
likely age greater than $3$\,Ga for volcanic resurfacing at the CBVC \citep{Shirley2013}, suggesting that the Th distribution exposed at the CBVC may provide a rare insight into the extent of fractionation and the distribution of such magmatic activity at this time in the Moon's evolution.
%
%

One drawback of gamma ray spectroscopy is the large spatial footprint of a gamma ray detector.  
When the LP-GRS was in an orbit $30$\,km above the lunar surface, the full-width at half-maximum (FWHM) of the detector's footprint was $\sim 45$\,km \citep{Lawrence2003}. 
Additional statistical analysis is
therefore
required to extract information about the chemistry of sites as small as the CBVC.
%
%
In this paper we use the pixon method \citep{PP93} to remove blurring caused by the large detector footprint and enhance the spatial resolution of gamma ray data (by a factor of $1.5-2$ compared with other image reconstruction techniques; \citealt{Lawrence2007}), in a way that is robust to noise.
%
%
This allows us to test the prevailing hypothesis regarding the distribution of Th around the CBVC --- that it is all contained within the high albedo region \citep{Jolliff2011}. Under that assumption, the raw counts from the LP-GRS data imply a Th concentration within the feature of $\sim 40\!-\!55$\,ppm \citep{Lawrence2003}, which is important because only one known lunar rock type has such high concentrations of Th, namely granite/felsite \citep{Jolliff98}.

%
%
\subsection{Lunar volcanism}
Basaltic volcanism was once common on the Moon and is responsible for the lunar maria that cover 17\% of the lunar surface \citep{Head1976}, mostly filling the near-side basins.  Evidence for basaltic, non-mare volcanism is most evident in the dark glasses that are distributed across the lunar surface, which are thought to be the product of basaltic pyroclastic eruptions.  

Silicic, non-mare volcanism is much less common, observed at only a handful of locations including Hansteen alpha \citep{Hawke2003}, Mairan hills \citep{Ashley2013}, Lassell Massif \citep{Hagerty2006,Glotch2011}, the Gruithuisen Domes \citep{Chevrel99} and Compton-Belkovich \citep{Jolliff2011}.  These silicic volcanic constructs are steep-sided, with widths of a few km and heights greater than 1\,km.
%
%
Their morphology, enhanced Th concentrations and Christiansen features all imply that these units are the result of evolved, silicic volcanism \citep{Hagerty2006,Glotch2010}. All of these units are located within the Procellarum KREEP Terrane, except the CBVC, which is on the lunar farside.  One explanation of the origin of the silicic domes is that 
%
%
they are formed by the eruption of magma that is produced when ascending diapirs of basaltic magma stall at and underplate the base of the crust, causing it to partially remelt; the resulting melt is more silicic than the original basalt, and is enriched in incompatible elements and phases \citep{Head2000}.
%
%
Alternatively the basaltic magma might stall at the base of the megaregolith, then undergo fractionation to produce a more evolved, silicic magma \citep{Jolliff2011LPI}, which subsequently erupts.
%
%

Silicic, non-mare volcanic centres have previously been assumed to be similar in nature to terrestrial rhyolite domes \citep{Hagerty2006}, which are erupted extrusively. \citet{Jolliff2011} suggest that pyroclastic material may have been distributed over distances of a few kilometres from the CBVC but, to our knowledge, no evidence has previously been presented for lunar volcanism that is both pyroclastic and silicic.

%
%
\subsection{Th rich minerals at the CBVC} 
The association of high Th concentrations in lunar samples with granite or felsite is clear, with Th concentrations of granitic samples generally falling in the range $20-65$\,ppm \citep{Seddio2013}.  Granitic assemblages clearly form from highly differentiated melt compositions that are enriched in many of the incompatible trace elements, especially the large-ion-lithophile (LIL) elements.  Mafic evolved assemblages also occur in the lunar samples that exhibit LIL enrichment, such as alkali anorthosite and monzogabbro \citep{Jolliff98}; however, these assemblages do not contain as high Th and U concentrations as do some of the granitic samples.  Alkali anorthosites have Th concentrations as high as $40$\,ppm, but most have $< 20$\,ppm, and monzogabbro samples have Th concentrations as high as about $45$\,ppm \citep{Wieczorek2006}, but they have substantially higher FeO ($\sim10-16\,\mathrm{wt.}\%$) than is indicated for the CBVC by LP-GRS data \citep{Jolliff2011}.  KREEP basalts only contain up to about $15$\,ppm Th and FeO typically in excess of $10\,\mathrm{wt.}\%$. 

%
%

\section{Data}\label{sec:Data}

\subsection{LP-GRS Data}\label{ssec:Data}
\begin{figure*}[t]
\begin{center}
\includegraphics[width=\textwidth]{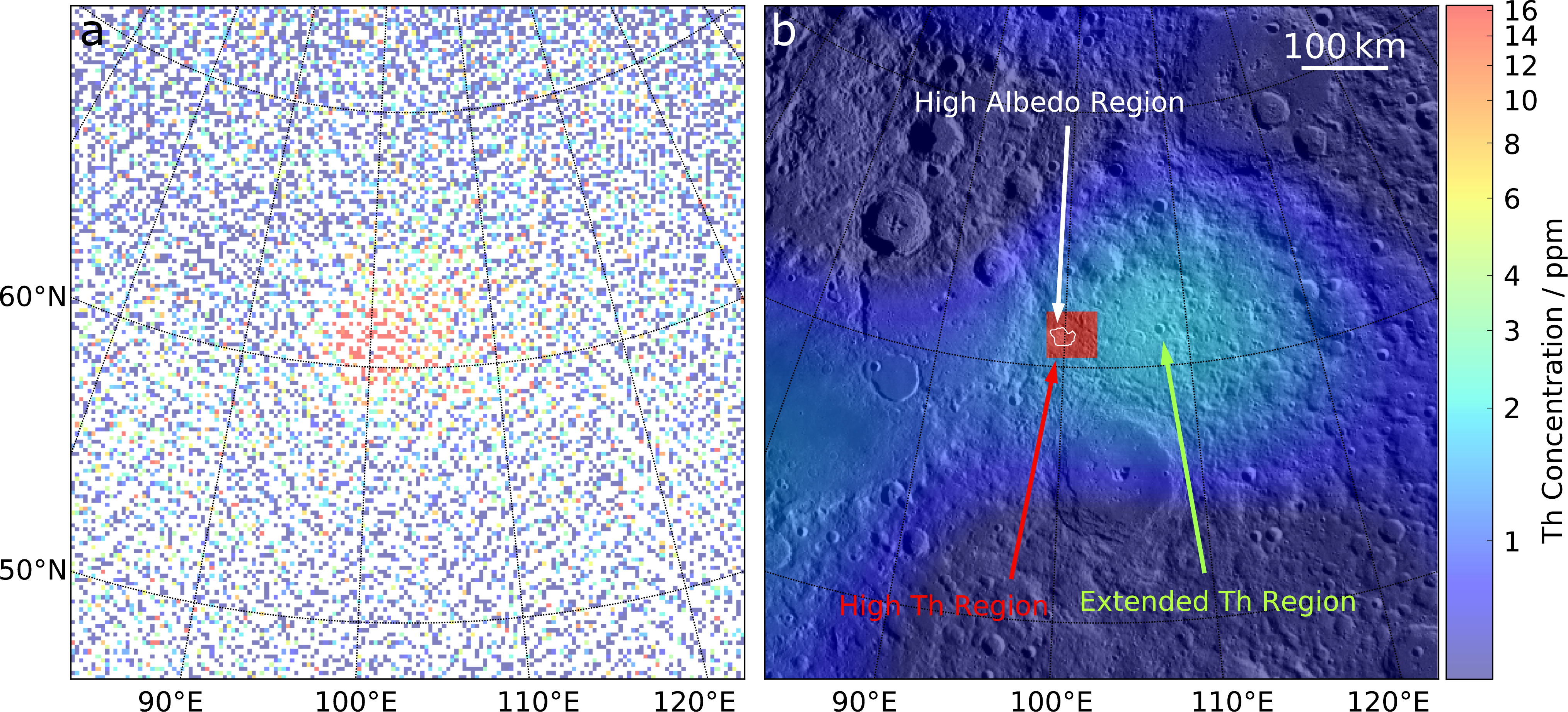}
\end{center}
\caption{{\bf a:} The raw LP-GRS data binned in the same pixellation that is used in the reconstruction. {\bf b:} The best-fitting, unblurred Th distribution in the vicinity of the CBVC. The size and position of the decoupled region are well constrained; the square shape shown is marginally more consistent with the data than a circle.
The high Th region is shown in red and occupies approximately $5$ times the area of
the high albedo feature, which is shown with the white contour. Underlayed is a WAC \citep{LROC2010} image of the area around the CBVC.}
\label{fig:recon}
\end{figure*}

Time-series LP-GRS observations from the $7$ months that the
Lunar Prospector spent at an altitude of $\sim 30$ km are used in this
work. 
%
%
Each observation accumulated a
gamma ray spectrum over an integration period of $\Delta t=32$\,s, giving 490952 
observations in total. 
%
%
The reduction of these data was described
by \cite{Lawrence2004}, with the counts in the Th decay line at $2.61$\,MeV 
being defined as the excess over a background value within the
$2.5-2.7$\,MeV range. Absolute Th abundances are determined following the
procedures used by \citet{Lawrence2003} and based in part on spectral unmixing work described in \citet{Prettyman2006}.
The typical count rate is $c=3.7$ gamma rays per second, with the
conversion to Th concentration in ppm being
\begin{equation}
[{\rm Th}]= 3.8912 c - 13.6584,
\end{equation}
from \citet{Lawrence2003}. A plot of this data is shown in Fig.~\ref{fig:recon}. The location with highest Th is coincident with the eastern edge of the albedo region, not its centre as may naively be expected.  This has led some to suggest a possible offset between the two data sets. If such exists, then it is sufficiently small that it should have no effect on the conclusions of this paper.

\subsection{Assumed instrumental properties}\label{ssec:Assump}
We begin with the LP-GRS PSF model by
\citet{Lawrence2004}, which is circular and has a FWHM of $\sim 45$\,km, but we modify this to
take into account the motion of the spacecraft. 
The detector moved at $\sim 1.6$\,km$\mathrm{s^{-1}}$ with
respect to the lunar surface, i.e.\,$\sim 51$\,km during the $32$\,s integration period.  
We convolve the circular PSF with a
line extended in the direction of motion of the spacecraft, with
length equal to the distance traveled by the spacecraft
during one observation. This produces an elliptical PSF that is elongated in the direction of the poles.

As the LP-GRS was a counting experiment, the number of Th decay gamma
rays received above a particular patch of lunar surface should follow
a Poisson distribution. However, during the data 
reduction process, corrections have been applied such that the reduced
count rates could have a somewhat different distribution. These
corrections compensate for temporal variations in the galactic cosmic ray
flux, the varying altitude and latitude of the spacecraft and the
detector dead-time (i.e.\ the interval after a detection in which
another cannot be registered) \citep{Lawrence2004}. Despite these
various non-negligible corrections, \citet{Lawrence2004} showed that
the noise on the Th-line data was surprisingly close to Poisson. In order that
the reconstruction method can appropriately weight each observation,
it is important to understand the statistical properties of the noise.

To gauge the effects of these corrections on the data we created a
mock set of data in which the noise was known to be Poisson and to
which each of the corrections (for temporal variations in the galactic
cosmic ray flux, the varying altitude and latitude of the spacecraft
and the detector dead-time, described in detail in
\citet{Lawrence2004}) were applied in turn. This mock data set was
created by:
\begin{enumerate}
\item undoing the corrections made to the Th-line LP-GRS time-series
  observations to find the measured number of counts in each
  observation; 
\item making a time-series of
Poisson random variables with means equal to the number of counts
found in the previous step;
\item applying the four corrections to create the mock time-series. 

\end{enumerate}
Once the mock time-series was created, the
statistics were tested by binning the data in $2\degree \times 2\degree$
pixels.  The expected scatter in observations within each of these
pixels is well approximated by a Gaussian with width
$\sigma=\sqrt{c\Delta t}$. This standard
deviation was compared with that measured from the repeat observations
of the same pixel. They were found to be in good agreement, with the
average ratio of the two differing from unity by less than $1\%$.  


This observed agreement results from the cancellation of the various corrections applied to the data: the correction of count rates to an altitude of $100$\,km typically decreases the corrected count rate by
$\sim 15\%$ whereas the normalisation of the background to the high
initial galactic cosmic ray flux increases the corrected count
rate by $\sim 10\%$ and correcting for deadtime increases it by $\sim
4\%$. Consequently, these factors approximately cancel and the variance of the corrected measurements can be accurately treated as
equal to the mean.

\section{Method}\label{sec:meth}

The aim of any image reconstruction is to arrive at the best estimate
of the true, underlying image ($I$) given an observed image (i.e. the
data $D$), an estimate of the instrumental point spread function
(PSF) or beam $B$, and perhaps some prior knowledge. An
observation can be described by the equation:
\begin{equation}
D({\bf x}) = (I*B)({\bf x}) + N({\bf x});
\end{equation}where ${\bf x}$ represents a pixel in the two-dimensional image, $N$
is the noise and $*$ represents the convolution operator.
As the distribution of the noise is known only statistically, there
is no hope of inverting this equation analytically in order to obtain
the unblurred truth, $I$. Instead we must resort to statistical
techniques that require a sound understanding
of the PSF, $B$, and the statistical properties of the noise, $N$, in order to find the ``inferred truth'', $\hat{I}$, that is most consistent with the data and is therefore our best guess at $I$.
A brief discussion of the data we use and our assumptions
about both $B$ and $N$ were
detailed in section~\ref{ssec:Assump}. The pixon image reconstruction
technique that we will use is described in section~\ref{ssec:Pix}.

\subsection{Pixon method}\label{ssec:Pix}
The technique we use to suppress noise and remove the effect of blurring with the PSF, thus arriving at the best estimate of the
underlying Th distribution,
is the pixon method \citep{PP93}, which has successfully been used
in a range of disciplines including medical imaging, IR and
X-ray astronomy [\citealt{P96} and references therein]. In addition, it
has recently been used to reconstruct remotely sensed neutron
\citep{Eke2009} and gamma ray data \citep{Lawrence2007} and has been
shown to give a spatial resolution $1.5$-$2$ times better than that of Janssen's
method in reconstructing planetary data sets \citep{Lawrence2007}.

The pixon method is an adaptive image reconstruction technique, in
which the reconstructed ``truth'' is described on a grid of pixons, where
a pixon is a collection of pixels whose shape and size is allowed to
vary. Thus areas of the image with a low signal to noise ratio are
described by a few large pixons, whereas regions of the data
containing more information are described by smaller pixons, giving the
reconstruction the freedom to vary on smaller scales. This method
is motivated by consideration of how best to maximize the posterior
conditional probability:
\begin{equation}\label{eqn:prob}
p(\hat{I},M|D) = \frac{p(D|\hat{I},M)p(\hat{I}|M)p(M)}{p(D)};
\end{equation}
where $\hat{I}$ is the inferred truth and $M$ is the model, which
describes the relationship between $\hat{I}$ and the data, including
%
%
the PSF and the basis in which the image is represented. As the data are
%
%
 already
taken, $p(D)$ is not affected by anything we can do and is therefore
constant. Additionally, to avoid bias, $p(M)$ is assumed to be uniform. This assumption
leaves two terms: the first, $p(D|\hat{I},M)$, is the likelihood of
the data given a particular inferred truth and model, which can be
calculated using a goodness-of-fit statistic, for example $\chi^2$ for
data with Gaussian errors, where
\begin{equation}\label{eqn:like}
p(D|\hat{I},M) = \exp(-0.5\chi^2),
\end{equation}
or directly from the probability density function in the case of data
with Poissonian errors. The second term, $p(\hat{I}|M)$, is the image prior -- the form of which can be deduced from counting
arguments. For an image made up of $n$ pixons and containing $C$ separate and indistinguishable
detections, the probability 
of observing $C_i$ detections in pixon $i$ is 
\begin{equation}\label{eqn:prior}
p(\hat{I}|M) = \frac{C!}{n^C \Pi^n_{i = 1}C_i!}.
\end{equation}
This prior is maximized, for a given number of pixons, by having the same
information content in each pixon i.e. $C_i = C/n$, for all $i$. The
image prior increases as fewer pixons are used, making it a
mathematical statement of Occam's razor 
%
%
and causing the image
reconstruction to yield an $\hat{I}$ that contains the least possible
structure whilst still being consistent with the data. 

Our implementation is based on the speedy pixon method described in
\citet{Eke2001} and \citet{Eke2009} but modified to allow a 'decoupled region' to vary independently from the region outside.  This will be used to reconstruct the high Th region immediately around the CBVC (a thorough description of the implementation is given in Appendix~\ref{sec:details}).

\section{Results}\label{sec:res}

We vary the size, shape, position and Th content of the decoupled region to determine the optimum reconstruction for the Th data in the vicinity of the CBVC. The resulting, unblurred image is shown in Fig.~\ref{fig:recon}. A high Th region that is $50\,\mathrm{km} \times 70$\,km, i.e. larger than the high albedo area, and a more extended lower concentration Th zone are both required by the data.
In this section, we will discuss the statistical significance of these features.
Our discussion divides naturally into considerations of the high Th region in the vicinity of the CBVC and the more extended spatial distribution of Th.
We consider the geological implications of our results in section~\ref{sec:imp}. 


\begin{figure}[t]
\begin{center}
\includegraphics[width=\columnwidth]{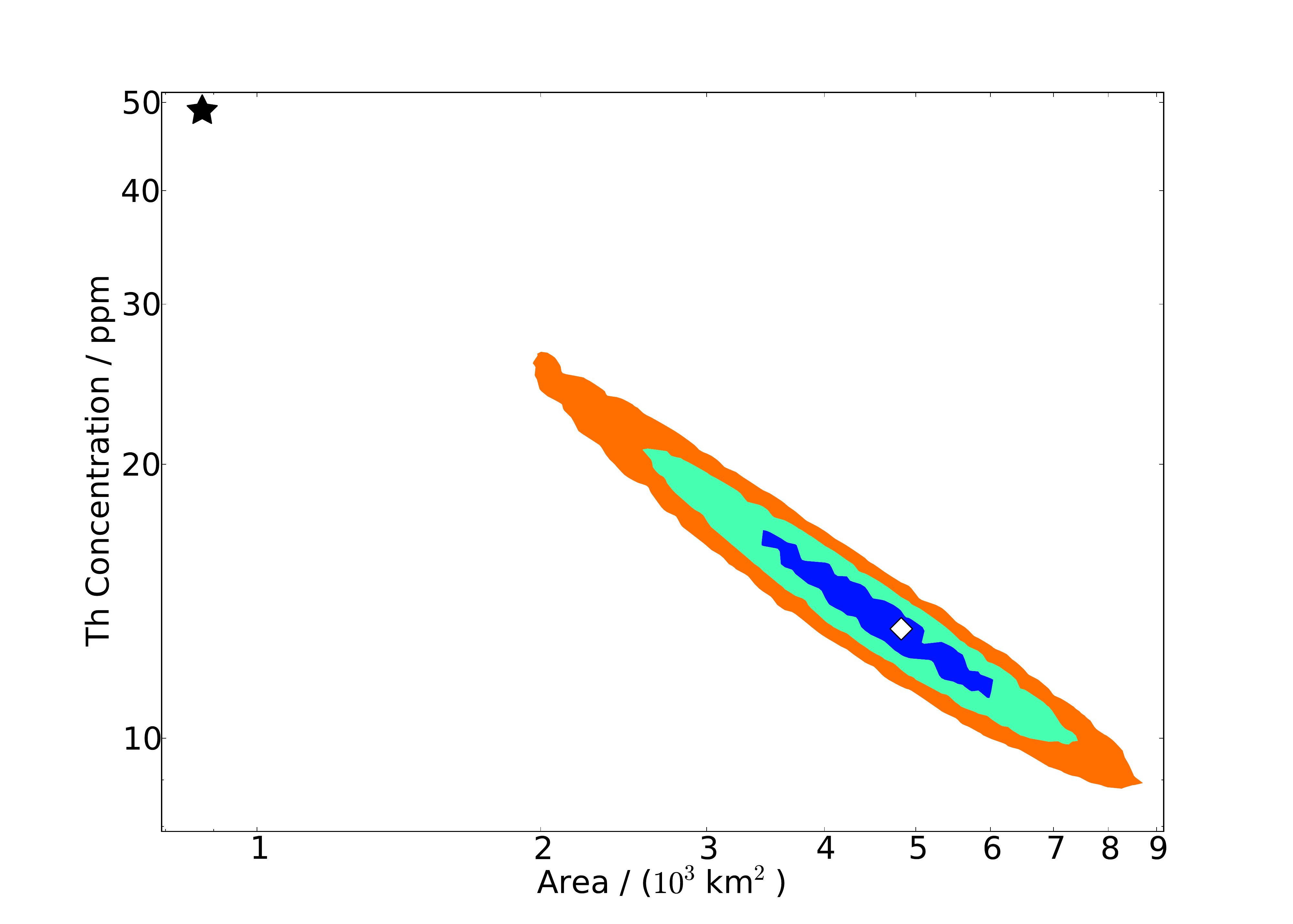}
\end{center}
\caption{Constraints on the size of the high Th region in reconstructed images, and the Th concentration inside that region.
The white diamond shows the best-fit reconstruction with minimum $\chi^2$. Coloured regions enclose 68\%/95\%/99.7\% confidence limits, determined using $\Delta\chi^2=\chi^2-\chi^2_\mathrm{min}$. 
The black star shows the optimum reconstruction under the assumption that the decoupled region coincides with the high albedo feature. 
}
\label{fig:chi}
\end{figure}

\subsection{The high Th region}\label{ssec:thin}

\citet{Lawrence2003} considered the Th excess at the CBVC to be localized within the $\sim 25\,\mathrm{km} \times 35$\,km high albedo region, at an abundance of $\sim 50$\,ppm. 
We have tested this hypothesis by examining a set of reconstructions that spread the Th excess across different sized and shaped high Th regions. While the rectangular high Th region shown in Fig.~\ref{fig:recon} is slightly favoured over a circular one, the difference is not large. Thus, to reduce the size of the parameter space to be tested, a circular high Th region centred on the middle of the high albedo region is adopted when considering the effects of changing the area of the high-Th region (any $< 10$\,km offset of the LP-GRS would shift this region by at most two pixels, an accuracy to which we are not sensitive).  This choice leaves only two free parameters; the area of the high-Th region and the Th concentration within it. 
The relative merits of these different choices are quantitatively assessed using the misfit statistic
\begin{equation}
\chi^2 = \sum_{i = 1}^m{R}(i)^{2},
\end{equation}
where $R(i)$ represents the residual in pixel $i$ (see Appendix A), and the sum is over the $m$ pixels within $60$\,km of the centre of the CBVC (this region focuses on the area where the highest Th count rates are concentrated, and is necessarily broader than the instrumental PSF).
The misfit statistic is driven mainly by the size of the decoupled region, rather than its shape or precise location. 

The position of the black star in Fig.~\ref{fig:chi} indeed indicates a preferred concentration of $49$\,ppm if the Th distribution is constrained to be within the $25\,\mathrm{km} \times 35$\,km high albedo region.
However, such a concentrated distribution of Th is very strongly disfavoured. 
The LP-GRS data are
%
%
 much better fitted by reconstructions in which the Th is uniformly distributed over $3400$--$5900$\,km$^2$, corresponding to diameters of $\sim 65$--$87$\,km, at lower concentrations of $\sim 17$--$11$\,ppm. 
%
This area is approximately 5 times larger than the high albedo feature, but still only slightly bigger than the LP-GRS PSF, so our measurement of its area is necessarily imprecise.  We have assumed that the PSF in \citet{Lawrence2003} is correct. If it were in fact larger or smaller then the results would change quantitatively.  However, in order to claim that all of the Th excess detected on the surface is contained within the high albedo region would require the PSF FWHM to be nearly twice its accepted value, which is not consistent with the work done by \citet{Lawrence2003} that places errors on the size of the PSF of a few km.  Such a small uncertainty on the assumed PSF does not change our results appreciably.
%
%


\subsection{The extended Th region}\label{ssec:thout}

Outside the central high-Th region there are two regions with enhanced Th content, the first, to the WSW, has a Th content less than $2$\,ppm and is coincident with the eastern edge of Mare Humboldtianum and so is not directly related to the silicic CBVC. The second has a Th content up to $2.3$\,ppm and extends $\sim 300$\,km east from the CBVC.
 This feature is evident in Fig.\,\ref{fig:recon}, in the results from the forward modelling in \citet{Lawrence2003}, and in the raw data \citep{Jolliff2011}. We will refer to it as the extended Th region. 

As a simple check of our procedure, we compute the statistical significance of the 
excess counts in the $50\,\mathrm{km}
\times 50$\,km square region centered $120$\,km east of the CBVC, which is sufficiently distant from the high Th area to receive few counts as a result of the PSF blurring. 
This area has a $5\sigma$
excess in counts, strongly suggesting that the extended Th region to
the east of the CBVC is a statistically significant Th excess.

A more detailed proof of the statistical significance of the extended Th region is given in Appendix~\ref{sec:pval}.

\section{Implications for the origin of the Th distribution}\label{sec:imp}

%
%
The results in the previous section imply that the high Th
region is larger than the $\sim 25\,\mathrm{km} \times 35$\,km area of silicic composition identified in Diviner data \citep{Diviner2010} and the area of increased reflectance identified in the Wide and Narrow Angle Camera (WAC, NAC) imaging \citep{Jolliff2011}. This result might imply that the
Th was emplaced in the high albedo region and has subsequently
undergone lateral transport to produce the current distribution, or
that the process that placed the Th on the lunar surface itself
imprinted this extended distribution. 

Assuming that the
%
%
high Th material was initially emplaced within the CBVC via silicic volcanism, as proposed by \cite{Jolliff2011LPI}, then the high albedo region can be taken to trace the original extent of the Th on the surface -- leaving its subsequent transport to be explained. 
%
%
Another possibility is that the original Th distribution was
not coincident with the high albedo region and that the presence of
the regions with elevated Th contents outside the albedo feature was caused by 
pyroclastic eruptions at a time close to or at the formation of the volcanic
complex.
%
%
This hypothesis was proposed by \cite{Jolliff2011} to
explain the eastward extension of the Th distribution beyond the high albedo region over distances of $\sim 7$\,km.

\subsection{The sputtering of Th atoms}\label{ssec:sput}
Sputtering liberates atoms from the surface of the lunar
regolith, but most sputtered atoms have speeds greater than the
escape speed \citep{Wurz2007} so sputtering tends to remove material
altogether. However, the most
probable speed of a sputtered particle with mass $m$ is expected to scale with
$m^{-\frac{1}{2}}$
\citep{Wurz2007}. As Th atoms have $m=232$\,amu the typical speed of a
sputtered Th atom is $\sim 0.8$\,km$\mathrm{s^{-1}}$, considerably less 
than the escape speed from the Moon ($\sim 2.4$\,km$\mathrm{s^{-1}}$).
Using the model by \citet{Cassidy2005} for the distribution of polar
angles, $\alpha$, of the sputtered atoms, 
\begin{equation}
f(\alpha) \propto \cos(\alpha),
\end{equation}
and assuming that the azimuthal
angular distribution is uniform, we can find the lateral velocity of
sputtered atoms and hence the average distance travelled by the
atoms before they fall back to the lunar surface, $d$.  This is done
by averaging over polar angle the product of the time of flight and lateral velocity.
Assuming for simplicity that the orbit is
parabolic, which turns out to be sufficiently accurate,
\begin{align}
\langle d\rangle &= \int^{\frac{\pi}{2}}_{0}\frac{2 v^2}{g}
 \cos^2(\alpha)\sin(\alpha) d\alpha,\\ &= \frac{2 v^2}{3 g},
\end{align}
where $v$ denotes the most likely initial speed of the sputtered atoms and $g$ is the
acceleration due to gravity. Using the most likely value of $v =
0.8$\,km$\mathrm{s^{-1}}$ gives $d = 250$\,km.

The equations above characterise the average hop of a sputtered Th
atom. However, to find the impact that this process has on the
concentration of Th in the vicinity of the CBVC we also need the rate of
sputtering.  A rough estimate of this can be made by ignoring binding
energy variations and assuming that the
number of atoms of a particular species that are sputtered from the
regolith is proportional to the number density of atoms of that
species in the regolith. Taking the sputtered flux of
oxygen given in \citet{Wurz2007} (where an average solar wind ion flux of $4.5 \times 10^{12}$\,m$^{-2}$s$^{-1}$ is assumed) and an oxygen concentration of
$46\,\mathrm{wt.}\%$ \citep{Heiken1991lunar} in the lunar regolith, versus a
typical Th concentration from our reconstructions of $20$\,ppm,
we estimate the flux of sputtered Th atoms to be $\Phi_\mathrm{Th} \sim 4 \times 10^5\,
\mathrm{m}^{-2}\mathrm{s}^{-1}$.  The average time a Th atom would
spend on the surface before being sputtered is then given by
\begin{equation}
\tau_\mathrm{sputter} = \frac{n_\mathrm{Th}d_\mathrm{sputter}}{\Phi_\mathrm{Th}},
\end{equation}
where $n_\mathrm{Th}$ is the volume number density of Th atoms, which is
related simply to the Th concentration and regolith density, and $d_\mathrm{sputter}=1$\,nm is assumed to be
the depth of regolith susceptible to sputtering.  Using the values
from above gives $\tau_\mathrm{sputter} \sim 13$\,yr. 

The final step in the consideration of this process is to assess the
effect of the overturn of regolith on the concentration at the
surface.  We assume that the rate of overturn of regolith due to
gardening is constant and $\sim 0.5\,\mathrm{m/Ga}$ \citep{Horz1977}. If
this gardening occurs at a constant rate, then the time that
Th atoms spend on the surface available for sputtering is
$\tau_\mathrm{overturn} \sim \tau_\mathrm{sputter}/10$. This implies that a
conservative approximation, when trying to determine the maximum amount
of Th that could leave the CBVC this way, is to
consider the Th at the surface of the CBVC to be constantly
renewed by the overturn of regolith.
Therefore the effect of sputtering on the CBVC is that, every year,
approximately one in ten atoms are sputtered and move away from the CBVC with
a typical step size of $\sim 250\,$km.  The sputtered atoms that are
re-implanted in the regolith are then most likely gardened down and
never take another hop.  The net effect over the $\sim 3.5$\,Ga of the
CBVC's lifetime is that $\sim 10\%$ of the Th atoms in the top $\sim
1$\,m of regolith, the region accessible to the LP-GRS data, will have
left the CBVC and settled in the surrounding few hundred km.  This dispersal would increase the Th concentration in the area surrounding the CBVC
by considerably less than $1\%$, which is not enough to explain the
findings of section~\ref{sec:res}.

It should be noted that the above argument places an upper limit on
the effect of sputtering on the concentration at the CBVC because of
the assumptions that the overturn of regolith occurs continuously and
that material, once gardened from the surface, is randomly distributed
throughout the underlying regolith.  In practice neither of these
assumptions hold exactly. \citet{Gault1974} suggested that the main cause
of gardening is impacts of small meteorites and consequently it is only
the upper mm of regolith that is continuously reworked and regolith
deeper than $1$\,cm is rarely brought to the surface \citep{Horz1977}.
Additionally if overturn is due, primarily, to micrometeorites it is
best to think of overturn taking place to a depth of approximately a
$\mathrm{\mu}$m every kyr and not as a continuous process.  We would,
in this case, expect of order $0.1\%$ Th to be lost from the CBVC instead
of the $\sim 10\%$ found above. 


\subsection{Mechanical transport of Th-bearing regolith}\label{ssec:lat}
Meteorites impacting on the lunar surface cause lateral mixing of
regolith. When
the regolith is made up from two compositionally distinct components
this lateral transport can lead to a diffusion-like effect in which
the two regolith types are mixed mechanically. The bulk regolith
composition at any point is a weighted average of the two end
states. We have calculated the effect of this process on the Th
concentration at the CBVC using the model described in \citet{Li2005}
and \citet{Marcus1970} under the assumption that the CBVC was originally a
compositionally homogeneous, circular feature $35$\,km in diameter
surrounded by a uniform background.

The lateral transport model assumes the following power law
relationships for the 
number of craters, $T$, above a given crater diameter, $x_c$, and the ejecta
thickness, $\zeta$, with distance from the crater rim, $r$,
\begin{align}\label{eqn:power}
T &= F x_c^\gamma, \\ \zeta(r) &= R_0x_c^h\left(\frac{x_c}{2r}\right)^k
\end{align}
where $F$, $R_0$, $h$ and $k$ are constants. We
set the constants of the crater rim ejecta height ($R_0$ and $h$)
using the data from \citet{Arvidson1975} as these are thought to be
relevant for small impacts ($x_c < 1$\,km) and it is presumably the frequent,
smaller impacts that contribute most to the dispersal of high Th
material from the CBVC. We take $F = 2.5 \times 10^{-3}$\,km$^{-2}$ as is appropriate for late Imbrian ages \citep{Wilhelms1987}.  The only constraints we place on $k$ are those
theoretical limits suggested by \citet{Housen1983}, that $2.5 < k <
3.0$. Consequently the model requires $\gamma$ to lie in a certain range as it
must obey the condition
\begin{equation}
2/k < \gamma / (k + h).
\end{equation}

Using this model and assuming cratering to be a random
process allows one to derive a relationship between the total
thickness of regolith, Z, at a particular location, that originated at
least some given distance, $\rho$, away from that location. The
formalism does not, however, give a value of the ejecta thickness at a
particular point - only the characteristic function of the probability
density function (p.d.f.) of the total ejecta thickness,
$\phi_\rho(u)$:
\begin{align}
\phi_\rho(u) &= \exp{-\lambda |u|^{\alpha_\rho}[1 - i
\operatorname{sgn}(u)\tan(\pi\alpha / 2)]},\\ \lambda &=
\frac{\pi\gamma R_0^{\alpha_\rho}F}{2\alpha_\rho k(\gamma - 2 -
2h/k)(2\rho)^{\alpha_\rho k - 2}} \Gamma(1 -
\alpha_\rho)\cos(\pi\alpha_\rho / 2), \\\alpha_\rho &= \gamma / (k +
h),
\end{align}
where $\Gamma$ is the Gamma function, $i = \sqrt{-1}$ and
$\operatorname{sgn}(u)$ is $1$ if $u > 0$ and $-1$  if $u < 0$.
Obtaining the p.d.f. from this characteristic function
requires the use of a numerical integrator, and we follow \citet{Li2005}
in using the STABLE code \citep{Nolan1999}.

After the p.d.f. is obtained, the mode of the distribution is taken as
the value of $Z(\rho)$.  We combine these results with our model of
the CBVC (that it was initially a circular feature $35$\,km in diameter) to find, as a function of distance from the CBVC, what fraction
of the current regolith originated within the CBVC. The 
assumption is made that the regolith is well mixed and that the Th
detected by the LP-GRS can be related directly to the proportion of
ejecta at a particular point that originated from within the CBVC.

\begin{figure}
\begin{center}
\includegraphics[width=\columnwidth]{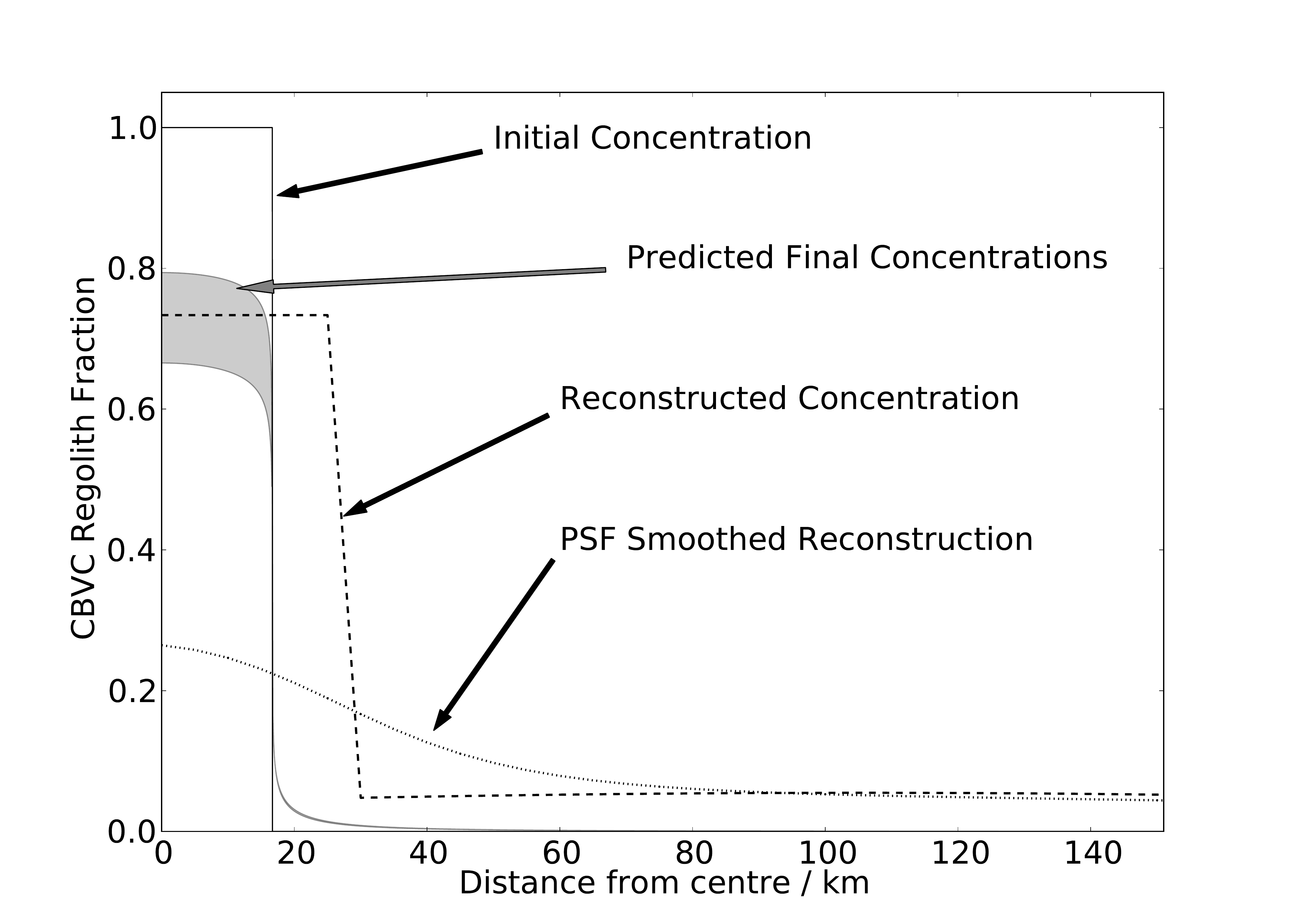}
\end{center}
\caption{Assuming no pyroclastic dispersal, the variation of the fraction of regolith that originated within the CBVC with distance from the centre of the CBVC. The solid black line shows the hypothetical initial CBVC Th distribution, at time zero.  The grey shaded area shows the range of solutions after 3.5\,Ga of modification by impact processes.  The dashed line is the present day reconstructed Th concentration (scaled so that it tends to zero at large distances and matches the model within the CBVC). The dotted line illustrates how the LP-GRS PSF suppresses the peak Th concentration being measured (scaled in line with the unsmoothed reconstruction).}
\label{fig:lateral}
\end{figure}

Fig.~\ref{fig:lateral} shows the initial assumed Th concentration profile with a black solid line.
The grey shaded region traces the variation, with distance, of the fraction of regolith that originated in
the CBVC $\sim 3.5$\,Ga earlier. A dashed line traces through the reconstructed 
Th map in the easterly direction, with
the vertical scale chosen so that the value tends to zero at large
distances and matches the model regolith fraction within the CBVC.
While repeated small impacts are
capable of moving some Th rich regolith away from the CBVC, it does not
increase the Th concentration in the $5$-$15$\,km around the CBVC to the
levels necessary to explain the difference between the high Th and
high albedo regions implied by the results of section~\ref{sec:res}.
The lateral transport model predicts that $\sim 25\%$ of the regolith
currently within the CBVC originated from outside this region and has
subsequently been redistributed into the CBVC by impacts. The effect of such transport would require that the measured Th concentration in the high Th region
should be correspondingly increased in order to infer the initial
concentration placed onto the surface by volcanism.
Also shown in Fig.~\ref{fig:lateral} is a curve showing the radial Th count rate variation in the reconstruction after it has been smoothed by the LP-GRS PSF, showing that the pixon method has increased the Th concentration in the high Th region by a factor of $\sim 3$ over that present in the blurred data.

In addition to the movement of regolith by impacts, one may hypothesise
that, as the CBVC is a topographically elevated feature, down slope
motion of regolith due to seismic shaking may be important in the lateral
transport of Th bearing regolith.  This is, however, not the case. The steepest slopes in the outer regions of the CBVC reach only $\sim\!10\degree$ from the horizontal, which, using the slope-distance relation from 
\citet{Houston1973} suggests lateral transport of $\sim 1$\,cm 
%
%
in 3.5\,Ga
due to seismically induced downslope motion.

\subsubsection{Effect of post-emplacement dispersal on Th content within the CBVC}
Although the processes described in sections~\ref{ssec:sput}--\ref{ssec:lat}
did not greatly affect the Th concentration in the region surrounding
the CBVC, lateral transport of regolith does have a significant effect
on the measured abundance of Th within the CBVC. The fraction of
regolith within the CBVC that originated there is between $\sim 0.65$
and $0.85$ (Fig.~\ref{fig:lateral}). This suggests that the Th
concentration when the CBVC was formed may have been $\sim 25-50\%$
greater than is detected today (approximating the surrounding regolith as having essentially a zero Th concentration). Consequently the Th concentration in
the high Th region inferred from the reconstruction
in section~\ref{ssec:thin} underestimates that present when the
material was emplaced onto the surface. As a result, the minerals that made up
the CBVC at emplacement would have contained $\sim 14$-$26$\,ppm Th
depending on the actual size of the feature and the parameters chosen
in equation~\ref{eqn:power}.  We have not included the effect of sputtering on the change in concentration as our estimate is an upper bound and we suspect that the true contribution of this process is somewhat lower than $10\%$; however, if we were to include it, then this would raise the upper limit on the allowed Th concentration to $29$\,ppm.

The above Th concentrations along with the low FeO content around the CBVC imply that the rock components that are most likely to be present at the CBVC are granite/felsite and alkali anorthosite or some combination.  In either case, the presence of alkali feldspar and a silica mineral (or a felsic glass) provides the best match for the LP-GRS Th and Fe data. 
 
\subsection{Lunar pyroclastic activity as a method of material transport}\label{ssec:pyro}
As has been shown in sections \ref{ssec:sput} -~\ref{ssec:lat} the
effects of post-emplacement processes to alter the distribution
around the CBVC are insufficient to explain the extent of the Th
distribution measured in the reconstructions in
section~\ref{ssec:thin}. Therefore the Th must have been initially emplaced
more widely than the high albedo region. 

We hypothesise, following
\cite{Jolliff2011}, that the mechanism of emplacement was pyroclastic eruption of a highly silicic kind
not readily evident elsewhere on the Moon.
%
%
Our results require dispersal over much greater distances of $\lesssim 300\,$km, than proposed by \cite{Jolliff2011}.
Repeated pyroclastic
eruptions from the many volcanic features in the CBVC could feasibly
give rise to the observed high Th regions that extend beyond the high
albedo feature.  The upper limit for ejecta distance for primitive pyroclastic eruption on the Moon is $350\,$km according to \citet{Wilson2003}. One would expect that, as a melt evolves and becomes more concentrated in the volatile species that drives eruption, the ejection velocity would increase, implying that the range observed in the reconstructions is reasonable.

The total volume of material ejected from the CBVC that has given rise to the broad extended Th region to the east can be calculated, assuming that the ejecta had the same Th concentration as the CBVC in the reconstruction.  The ejecta depth is
\begin{equation}
t = \frac{[\mathrm{Th}] - [\mathrm{Th}]_{\mathrm{background}}}{[\mathrm{Th}]_{\mathrm{ejecta}} - [\mathrm{Th}]_{\mathrm{background}}},
\end{equation}
where $[\mathrm{Th}]$ is the Th concentration at a given point, $[\mathrm{Th}]_{\mathrm{ejecta}}$ is the assumed Th concentration of the pyroclastic deposits and $[\mathrm{Th}]_{\mathrm{background}}$ is the Th concentration in the surrounding regolith.  Integrating the ejecta depth over the feature gives an estimate of the total ejecta volume of $8\,\mathrm{km^3}$.

One may expect that the silicic material laid down during these pyroclastic events would be detected by Diviner; however, no spatial extension of the polymerized Christiansen feature position is seen much beyond the extent of the high albedo region \citep{Jolliff2011}.  This is readily explained since Diviner (and visible imaging) is sensitive only to the very surface composition whereas the LP-GRS is sensitive to a metre or so of depth.  During 3.5 Ga of regolith gardening the silica emplaced by pyroclastic deposition, could have been mixed into the upper meter of regolith and effectively obscured from Diviner, whereas the Th signal would remain visible to the deeper-sensing GRS. This same argument applies to the non-detection of volatile rich material outside the CBVC in M$^3$ data \citep{Bhattacharya2013,Petro2013}. 

\begin{figure}[t]
\begin{center}
\includegraphics[width=\columnwidth]{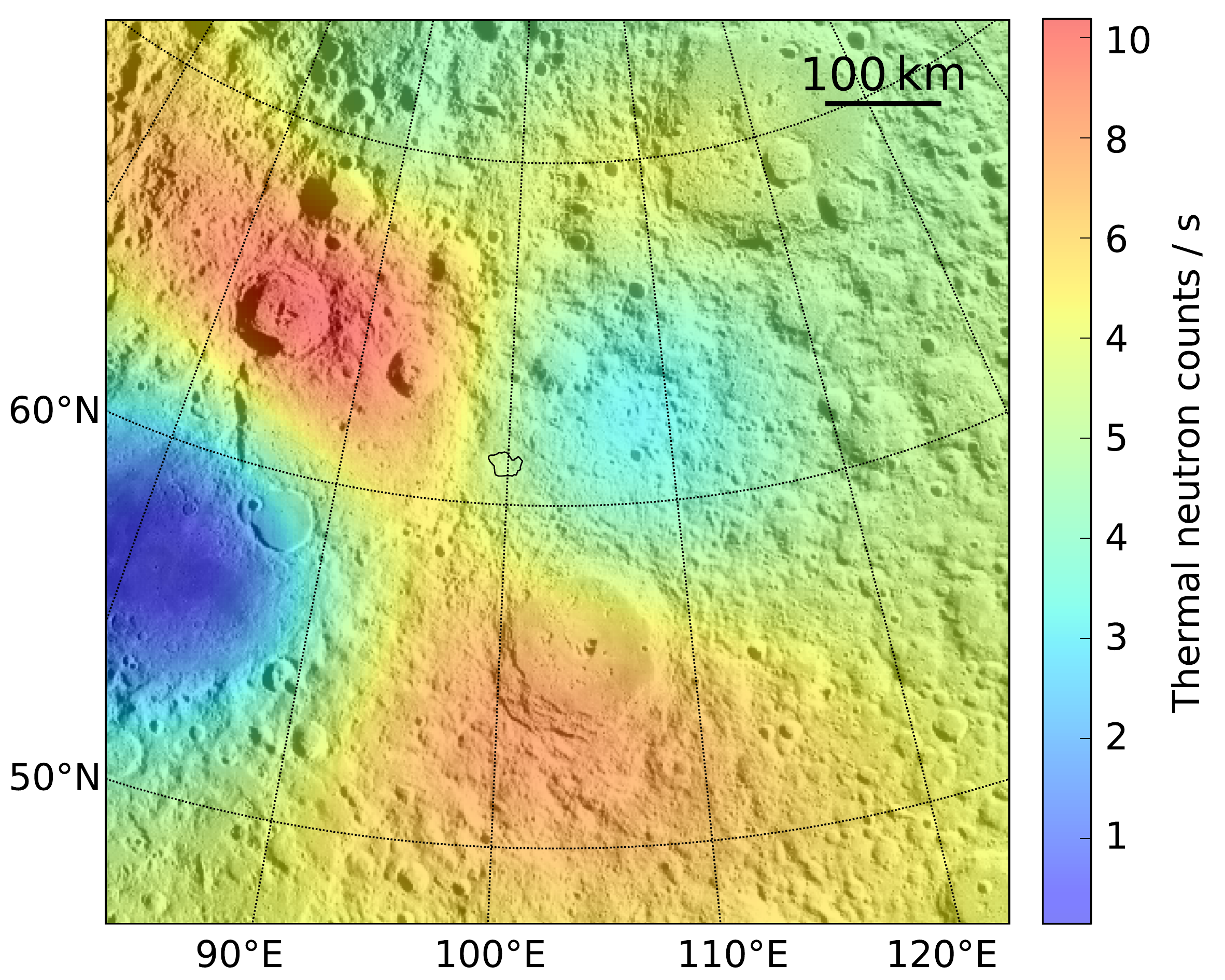}
\end{center}
\caption{The thermal and epithermal neutron count rate distribution in the vicinity of the CBVC, unblurred using a pixon reconstruction. The black contour shows the outline of the high albedo region identified by \citet{Jolliff2011}. Underlayed is a WAC image of the area around the CBVC.}
\label{fig:neutron}
\end{figure}

Nonetheless we would expect to see evidence of compositional difference in the extended Th region in data sets that are sensitive to some depth below the surface.  The LP neutron data probes the top ~1\,m of regolith.  A pixon reconstruction of the combined count rate of thermal and epithermal neutrons, measured by the LP spacecraft\citep{Feldman2000}, is shown in Fig.~\ref{fig:neutron}.  It is  clear that the extended Th region to the east of the CBVC once again shows up as compositionally distinct from its surroundings.

\section{Viability of lunar silicic pyroclastic volcanism}
Typical lunar pyroclastic eruptions are driven by primitive magmas and give rise to dark coloured deposits \citep{Head2002,Wilson2003}.
%
%
The pyroclastic deposits we propose as the cause of the extension of the high Th region would be expected to have a high albedo because of their silicic (low-Fe) composition --- akin to rhyolitic ash --- which gives rise to a light-colored assemblage of silica, alkali feldspar minerals and/or felsic glass. Smaller clast size also promotes higher albedo due to enhanced light scattering.

On Earth, explosive silicic volcanism produces abundant ash ($< 2$\,mm) and fine ash ($< 64$\,$\mu$m), which basaltic volcanism rarely does. Close to the vent, material is ejected as a jet, but the high surface area to volume ratio of the ash promotes rapid heat exchange with entrained atmosphere, producing buoyant, lofting plumes that may ascend tens of kilometres before attaining neutral buoyancy and spreading laterally \citep{sparks1982explosive}. Winds then dominate the subsequent dispersal of the ash, which may circle the globe in the case of the largest eruptions. By contrast on planetary bodies with negligible atmosphere such as the Moon, buoyant lofting is impossible. Instead, ejected particles will follow essentially ballistic trajectories, modified to some extent by particle--particle collisions. 

We can calculate the maximum distance ejecta might be expected to travel in a lunar pyroclastic eruption using the 1-dimensional gas flow equations. We treat the flow as a single-phase perfect gas and assume that ash particles will be accelerated to similar speeds to the gas.
%
%
This is the standard approach in planetary science and has been verified experimentally \citep{kieffer1984laboratory}. Whilst more sophisticated multi-phase modelling is possible, our simple approach is sufficient to obtain rough estimates for attainable velocities.


We consider a gas of density $\rho$, temperature $T$, and velocity $u$
moving steadily and essentially one-dimensionally along a conduit of
cross sectional area $A$. For an ideal gas with adiabatic index
$\gamma$, gas constant $R_g$, and specific heat at constant pressure
$c_p$, we have the following relations between inlet (subscript 0) and outlet (subscript 1)
conditions:
\begin{eqnarray}
  \frac{u_1}{u_0}     &=& \frac{A_0}{A_1}\left(\frac{T_0}{T_1}\right)^{\frac{1}{\gamma-1}}\\
  \frac{\rho_1}{\rho_0} &=& \left(\frac{T_1}{T_0}\right)^{\frac{1}{\gamma-1}}\\
  \frac{p_1}{p_0}     &=& \left(\frac{T_1}{T_0}\right)^{\frac{\gamma}{\gamma-1}},
\end{eqnarray}
where $2c_p(T_0-T_1) =u_1^2-u_0^2$. Eliminating the cross
sectional area, these equations can be rearranged to give the outflow
velocity
\begin{equation}
  u_1=\sqrt{u_0^2 +2c_pT_0\left[ 1- \left(\frac{p_1}{p_0}\right)^{\frac{\gamma-1}{\gamma}}\right]}.
\end{equation}
Expansion is limited by the condition that $p_1$ must
be greater than or equal to the atmospheric pressure. Hence, if atmospheric
pressure is significant, then this limits the exhaust velocity and leads to
a complicated shock structure \citep{kieffer1984laboratory}. On the
Moon however, $p_1/p_0$, is negligible and, assuming $u_1\gg u_0$, we have
a rough estimate for the largest attainable velocity
\begin{equation}
  u_1=\sqrt{2c_pT_0}.
\end{equation}
This has the very simple physical interpretation that all the thermal energy in the gas molecules is converted to linear kinetic energy; the correct form for an imperfect gas is
\begin{equation}\label{eqn:u1}
  u_1=\sqrt{2\int^{T_0}_0c_p\,dT}.
\end{equation}
For a temperature of 1100\,K (a typical eruption temperature for a silicic magma on Earth)
%
%
this gives a velocity of
1\,430\,m\,s$^{-1}$, where it has been taken that the gas will be predominantly carbon monoxide, which is produced in oxidation-reduction reactions between native
carbon and metal oxides on nearing the surface \citep{Fogel95}.
%
%
%
%
This can be shown to be substantially greater than the launch speed necessary to emplace
debris ballistically over $300\,\mathrm{km}$, as is required by the reconstructions at the CBVC. The maximum range, measured along a
great circle, for an object launched at speed $v$ on an airless body
of mass $M$ and radius $R$ is
\begin{equation}
  x=2 R \sin^{-1}\left(\frac{Rv^2}{2GM-Rv^2}\right).
\end{equation}
For the Moon, where
$g=1.62\,\mathrm{m\,s}^{-2}$ and $R=1.74\times10^6$\,m, a speed of at least $669$\,m\,s$^{-1}$ is
necessary for a projectile to
travel $x=300\, \mathrm{km}$.

From equation (\ref{eqn:u1}), we calculate that a starting gas temperature of only 390\,K is enough to achieve the required velocity of 669\,m\,s$^{-1}$ under perfect conditions. Of course many factors, including non-optimal vent orientation,  will reduce these ideal ejection velocities but this simple
calculation shows that it is straightforward for volcanic plumes on
the Moon to eject material many hundreds of kilometres.
%
%

%
%
%
%
On Earth, volcanic conduits of all types may be inclined, and the vents are commonly asymmetric \citep{Wood1980,Folch2005,Castro2013}.
%
%
For basaltic pyroclastic eruptions, which typically emplace ballistically, this may cause asymmetry in the
distribution of pyroclastic material. For silicic eruptions, the rapid formation of a buoyant plume tends to disguise and overprint the effects of conduit inclination and vent asymmetry: the plume takes the ash straight up and wind is then dominant. In the absence of an atmosphere, silicic eruptions, too, would emplace ballistically, thus an inclined conduit would give rise to an asymmetric deposit. Furthermore, \cite{Jolliff2011} note the presence of arcuate features in the topography of the CBVC that are suggestive of the collapse of volcanic edifices. Such collapses have been known to trigger directed, lateral blasts in silicic volcanoes on Earth, the 1980 eruption of Mt.\ St.\ Helens being the most famous example \citep{Kieffer1981}.
%
%
It therefore seems plausible that the asymmetric distribution of pyroclastic material at the CBVC might have been caused by eruption from an eastward-inclined conduit, or from an asymmetric vent open to the east, perhaps as a result of the collapse of a volcanic edifice.
%
%

Evidence for both effusive (dome-forming) and explosive (pyroclastic) eruption is seen at the CBVC.  On Earth it is common for silicic volcanoes sometimes to erupt effusively, and sometimes explosively. This may be a consequence of variable differentiation of the melt, or variable composition of the magma at the point of formation \citep{Sides2014}, both of which may influence the viscosity and volatile content of the magma. %
%
The 2011 eruption of Cord\'on Caulle in Chile further demonstrated that silicic magma may simultaneously erupt effusively and explosively from a single vent \citep{Castro2013}. The variation in eruption style was inferred to result from variation in the path travelled by the magma during its ascent of the conduit, affecting its capacity to degas \citep{Castro2013}. An inclined conduit and asymmetric ash jetting --- quickly masked by the formation of a buoyant plume --- were also inferred for this eruption.  All of these features suggest that, were it on Earth, the CBVC would not be so unusual a volcanic feature.

\section{Conclusions} \label{sec:conc}
%
%
We have used the pixon image reconstruction method to
produce the highest 
resolution map of the Th distribution around the Compton-Belkovich
Volcanic Complex to date. 
%
%
This method largely removes the effect of the detector footprint from the Th map, in a way that is robust to noise present in the data.
A central excess of Th had been previously assumed to be coincident
with the $25\,\mathrm{km} \times 35$\,km high albedo region observed in LP-NAC/WAC imaging \citep{Lawrence2007,Jolliff2011}.
However, we have shown that the central Th excess likely extends $\sim\!65$--$87$\,km laterally.
The Th concentration in
this region would have been $\sim\!14$--$26$\,ppm at emplacement, with uncertainty driven by the precise current area and the amount
of external Th-poor regolith that has been mixed into the CBVC during the past 3.5\,Ga. 

We identify an additional Th feature (significant at $>5\sigma$), which extends $\sim\!300$\,km east of the CBVC at a Th concentration of $\sim\!2$\,ppm.
The data outside the CBVC are certainly not consistent with a uniform low-Th background.
The extended nature of the CBVC is not due to
processes that have acted since its origin (e.g. lateral transport of
regolith and sputtering) so must have been present when it was formed.

These silicic distributions of Th are consistent with a mixture of pyroclastic eruptions, to
distribute the Th widely, and effusive eruptions to produce the
observed volcanic domes and high albedo region.

\appendix

\counterwithin{figure}{section}
\section{Implementation of the pixon method}\label{sec:details}

The inferred Th count-rate map, $\hat{I}$, is based on a pseudoimage that is defined in the same $5$\ km pixel grid as the data. $\hat{I}$ is constructed by convolving this pseudoimage,
$H$, with a Gaussian kernel, $K$, whose width, $\delta$,
may vary across the image, i.e.
\begin{equation}
\hat{I}({\bf x}) = (K_{\delta({\bf x})}*H)({\bf x}).
\end{equation}
The local smoothing scale, or pixon width, $\delta({\bf x})$, is determined
for each pixel such that the information content is constant in each
pixon over the entire image (given by 
$\Upsilon({\bf x}) = (K_{\delta({\bf x})}*\upsilon)({\bf x})$, where
$\upsilon({\bf x})$ is the signal-to-noise ratio in pixel ${\bf x}$, which, using the fact that the data are Poisson distributed, is the square root of the total number of countsdetected in the pixel). 
In practice a finite set of pixon sizes is used as this keeps
the time of a single calculation of the misfit statistic down to
$\mathcal{O}({n_{\mathrm{pixels}}\log{n_{\mathrm{pixels}}}})$ (i.e. that of a
fast Fourier transform), where $n_{\mathrm{pixels}}$ is the number of
pixels in the image, whereas if the pixon size were allowed to vary
continuously then the time would be
$\mathcal{O}({n_{\mathrm{pixels}}^2\log{n_{\mathrm{pixels}}}})$. In order to
generate $\hat{I}$ we interpolate linearly between the images based on
the two pixon sizes closest to those required for each pixel.
\begin{figure}[h]
\begin{center}
\includegraphics[width=\columnwidth]{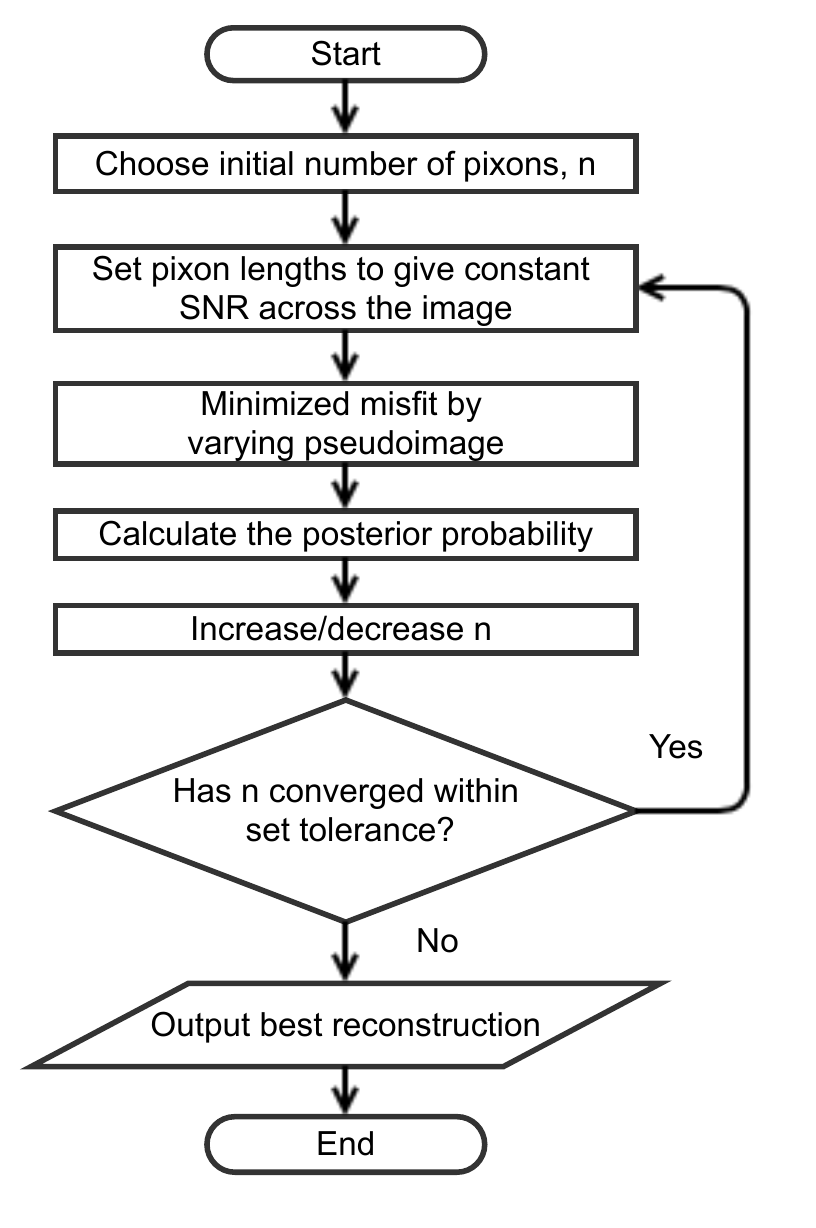}
\end{center}
\caption{A flow chart illustrating the algorithm used in the reconstructions.}
\label{fig:flow}
\end{figure}

The maximization of the posterior (equation ~\ref{eqn:prob}) is
done iteratively, with two stages in each iteration. Firstly, for a
given number of pixons, the pixon sizes as a function of position
%
%
are set so as to
maximise the image prior (equation ~\ref{eqn:prior}). Secondly, the 
%
%
values in the pseudoimage are adjusted to minimise the misfit statistic using the Polak-Ribi\`ere conjugate gradient minimization algorithm
\citep{NR}. 

The misfit statistic is derived from the reduced residuals between the
data and the blurred model, 
\begin{equation}\label{eqn:res}
R({\bf x}) = \frac{D({\bf x}) - (\hat{I}*B)({\bf x})}{\sigma({\bf x})},
\end{equation}
where $\sigma({\bf x})$ is the anticipated statistical noise in pixel
${\bf x}$.
Rather than using $\chi^2=\sum R^2$, we adopt $E_R$ from \citet{PP92} as the
misfit statistic. This statistic is defined as
\begin{equation}
E_R = \sum_{{\bf y} = 0}^mA_{R}({\bf y})^{2},
\end{equation}
where $A_R$ is the autocorrelation of the residuals, $A_R({\bf y})
= (R \otimes R)({\bf y})$ for a pixel separation, or lag, of ${\bf y}$. 
\begin{figure}[t]
\begin{center}
\includegraphics[width=\columnwidth]{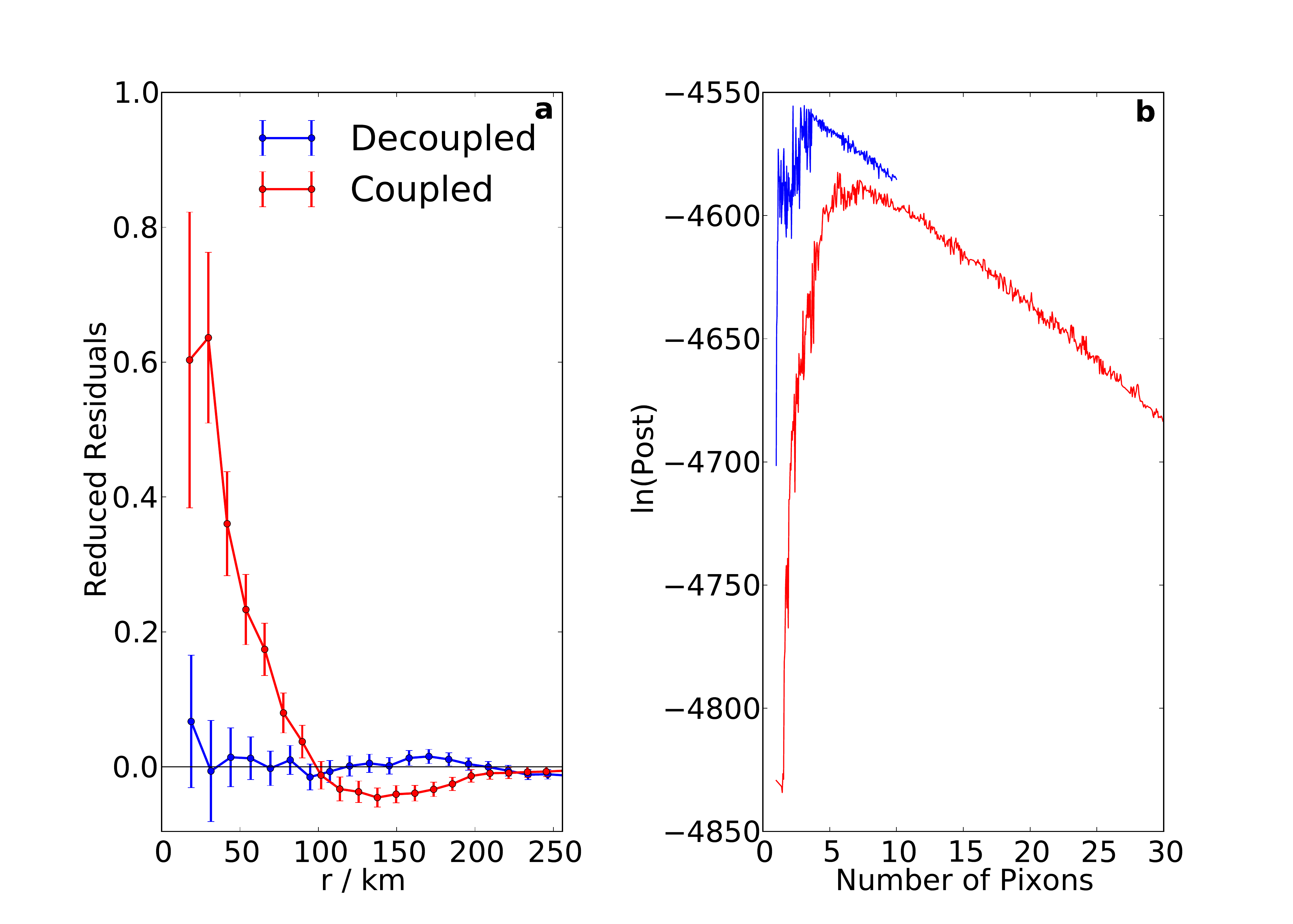}
\end{center}
\caption{{\bf a:} mean residuals as a function of radius from the centre of the CBVC.  
{\bf b:} the posterior probability of the reconstruction as a function of the number of pixons used.
In both panels, red (blue) points show the performance without (with) decoupling.
The corresponding, decoupled, inferred 2D Th abundance is shown in Fig.~\ref{fig:recon}.}
\label{fig:radres}
\end{figure}
The benefit of minimizing $E_R$, over the more
conventional $\chi^2$, is that doing so suppresses spatial
correlations in the residuals, preventing spurious features
being formed by the reconstruction process. \citet{PP92}
recommend that the autocorrelation terms defining $E_R$ should be those corresponding to pixel separations smaller than the instrumental
PSF. For our well-sampled PSF with $5$\,km square pixels, 
%
%
this means
many different pixel 
lags. However, for the reconstructions we have attempted
there 
is negligible
difference between those including different
numbers of pixel lags. We therefore use only the four distinct terms
with adjacent pixels (including diagonally adjacent pixels) to speed up the
computation.

%
%
In subsequent iterations, the number of pixons is varied in
order to maximize the posterior probability, or in practice its
logarithm, which by combining
equations~(\ref{eqn:prob}-\ref{eqn:prior}) and using Stirling's
approximation gives 
\begin{equation}\label{eq:post}
\ln(p(\hat{I},M|D)) = 0.5(\ln(N) + (1 - n) \ln(2\pi) \\ - n \ln(N/n) -\chi^2).
\end{equation}

\subsection{Decoupling}\label{sec:Dec}

The pseudoimage smoothing in the basic pixon method described in the previous section makes this algorithm unable to produce sharp boundaries in a reconstruction.  If such boundaries are demanded by the data then the residuals will be large, indicating that the reconstructed image represents a bad fit to the data. For the basic pixon reconstruction of the LP-GRS Th data in the region of the CBVC, such a problem occurs. The red line in the left panel of Fig.~\ref{fig:radres} shows the radial variation of the residuals from the point with highest Th, when using a basic pixon reconstruction. The positive residuals at small separations reflect an underestimation of the central reconstructed Th abundance resulting from the pixons oversmoothing this part of the reconstruction.

To incorporate very high spatial resolution features in the reconstruction, where we have prior information that they may exist (in this case provided by \citet{Jolliff2011} in the form of Diviner data and optical imaging) we exploit a technique described by \citet{Eke2009}.
Within a region of the image marked out by high residuals and prior information, which we shall call the ``decoupled region'', we introduce a separate pseudoimage that does not affect the reconstruction outside the decoupled region (and vice versa).  For the reconstructions in this paper, the decoupled region is effectively the high Th region including the CBVC.

Using a decoupled rectangular area of size $50\,\mathrm{km} \times 70$\,km centred on the CBVC, leads to the results as shown with blue curves in Fig.~\ref{fig:radres}. In addition to removing the non-zero radially-averaged residuals, as shown in the left hand panel, 
the posterior probability in the right-hand panel also reflects how decoupling dramatically improves the reconstruction. This demonstrates that the new technique gives rise to an inferred truth, i.e. reconstruction, that is more consistent with the data.

Unfortunately, the posterior probability curves are too noisy for their maxima to be easily located in an automated way.
Indeed, this noise is further increased when using decoupling, because the image optimization algorithm converges to fits in which the total Th abundance in the decoupled region varies.  We therefore chose to fix the count rate within the decoupled region, then scan a range of possible count rates for each step in the posterior maximization.  We can, in any case,  say little about how the count rate varies within the decoupled region, because the instrumental PSF is comparable to the size of the decoupled region. Nonetheless, we have explicitly tested distributions of Th within the decoupled region that decrease with distance like a Gaussian, and these are slightly disfavoured by the data.

\section{Statistical significance of the extended Th region}\label{sec:pval}

Any noisy image reconstruction may contain spurious features that are not demanded by the data. 
We shall therefore assess the statistical significance of this extended enhanced Th region, by determining the probability that a similar excess would have been reconstructed by chance, even if it had not actually been present in the underlying map.

\begin{figure}[t]
\begin{center}
\includegraphics[width=\columnwidth]{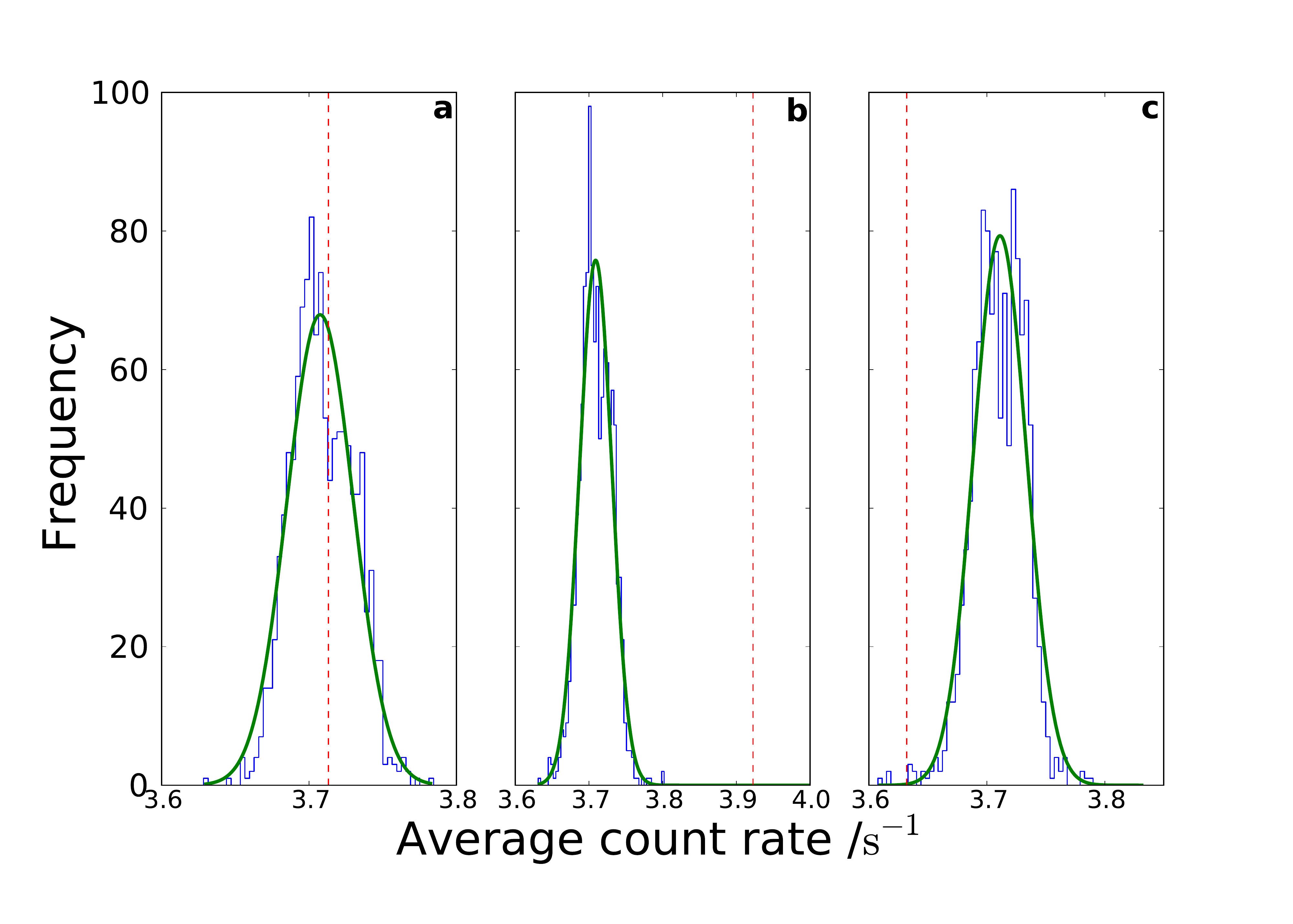}
\end{center}
\caption{Comparison of the reconstructed Th count rate from real LP-GRS data, and from mock data assuming no extended Th features, 
in 3 different map pixels ({\bf a}, {\bf b} and {\bf c} marked in Fig.\,\ref{fig:ps}).
The distributions from the mock
data sets are the blue histograms, with best-fitting Gaussians
shown with green curves. In each panel the vertical dotted red line is the
count rate in the corresponding pixels in the reconstruction of the
LP-GRS data. 
The probability of 
obtaining less than the predicted count rate can be found by
integrating the fit to the count rate distribution up to the red dashed
line.}
\label{fig:phist}
\end{figure}

\begin{figure}[t]
\begin{center}
\includegraphics[width=\columnwidth]{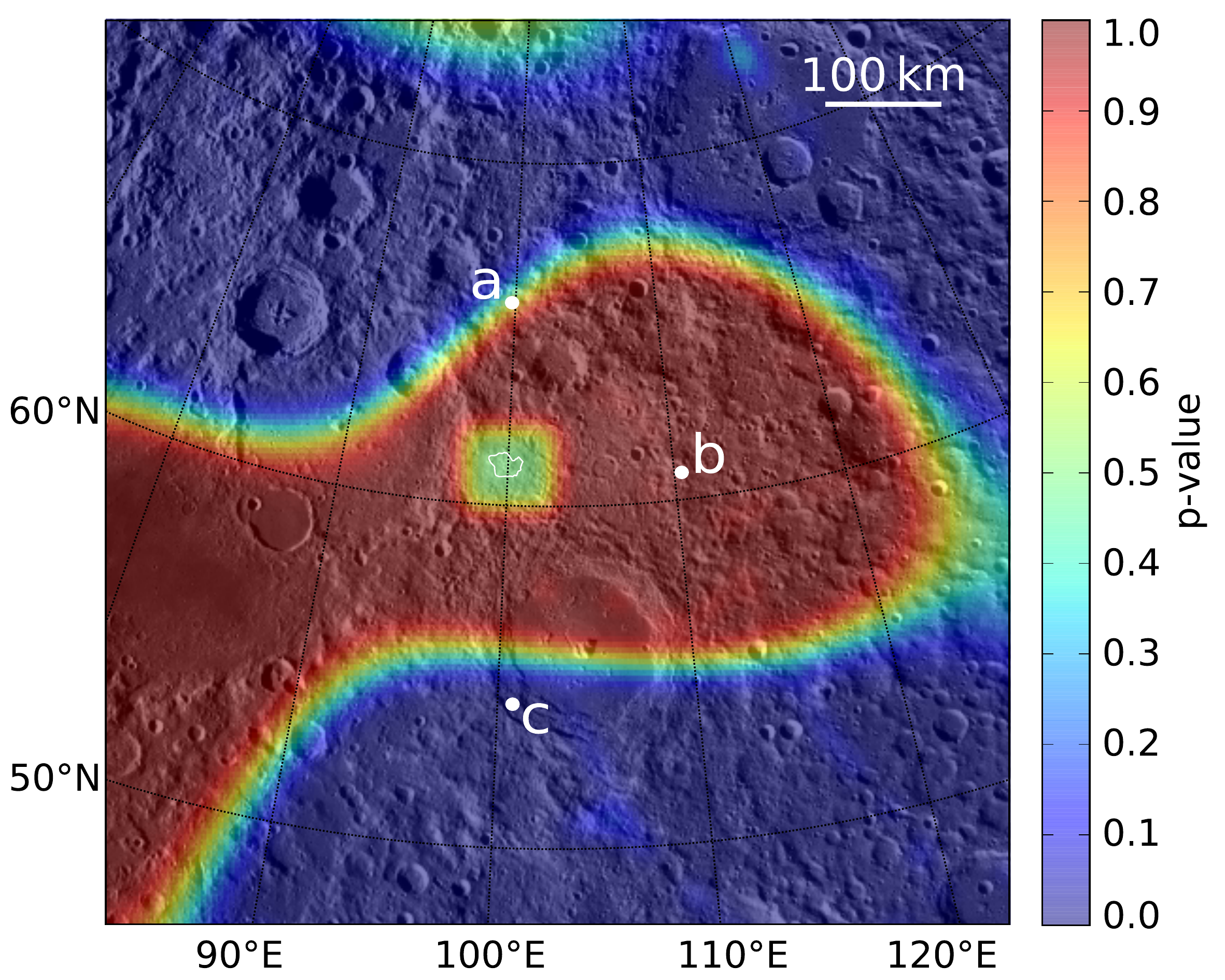}
\end{center}
\caption{The probability that the predicted count
rate is less than that obtained from the reconstruction of the LP-GRS
Th-line data set assuming the null hypothesis to be true in each pixel. The results 
have been smoothed with a Gaussian kernel with FWHM equal to that of the PSF
in order to suppress noise. Underlayed is a WAC image of the area around the CBVC.  The points labelled a, b and c mark the locations detailed in Fig.~\ref{fig:phist}}
\label{fig:ps}
\end{figure}

We simulated $1200$ mock data sets of a model with no extended Th excess; the model Th concentration is high inside the decoupled region, and low outside (equal to the mean of pixels outside the decoupled region in Fig.~\ref{fig:recon}).
%
%
Each mock data set is created by blurring this model map with the instrumental PSF, then taking  a different noisy realisation to give the observed gamma-ray count rates.
These simulated data are then reconstructed,
producing $1200$ Th count rate values in each map pixel, from which we can work out the probability of any false positive reconstructed excess being found in the actual LP-GRS Th reconstruction.
Fig.~\ref{fig:phist} shows the distributions of simulated count rates for the three pixels labelled in Fig.~\ref{fig:ps}. 
Also shown are the 
best-fitting Gaussian curves through the mock count rate
distributions. These curves describe the results well and allow us to
extrapolate to probabilities outside the range accessible with only
$1200$ samples. Pixel 1 lies in a region where the mean
reconstructed model background count rate matches that in the
reconstruction of the LP-GRS data, whereas the LP-GRS data have
%
%
 a higher count rate in pixel 2 and lower one in pixel 3.

In each pixel, one can calculate the probability that the
reconstructed mock Th count rate is lower than that obtained from the
reconstruction of the LP-GRS data. This allows us to test the null
hypothesis underpinning the mocks, namely that the excess Th is
concentrated entirely into the high Th region. Doing so
yields the results shown in Fig.~\ref{fig:ps}. 
Regions of Fig.~\ref{fig:ps} with probabilities close to either $0$ or
$1$ should be considered unlikely.
The existence of such regions suggests
that there are significant variations in the Th concentration outside
the CBVC. 
One statistically
significant extended zone of excess Th reaches up to $\sim
300$\,km east of the CBVC.
%
%

\begin{acknowledgments}
LP-GRS and LRO data are available from NASA's Planetary Data System at
http://pds-geosciences.wustl.edu. The program STABLE is
available from J. P. Nolan's website:
academic2.american.edu/$\sim$jpnolan. The work of the Diviner and LROC teams are gratefully acknowledged.

JTW and VRE are supported by the Science and Technology
Facilities Council [grant numbers ST/K501979/1, ST/L00075X/1].
RJM is supported by a Royal Society University Research Fellowship.
This work used the DiRAC Data Centric system at Durham University,
operated by the Institute for Computational Cosmology on behalf of the
STFC DiRAC HPC Facility (www.dirac.ac.uk). This equipment was funded
by BIS National E-infrastructure capital grant ST/K00042X/1, STFC
capital grant ST/H008519/1, and STFC DiRAC Operations grant
ST/K003267/1 and Durham University. DiRAC is part of the UK national
E-Infrastructure.

We would like to thank two anonymous reviewers for their thoughtful comments that led to a much improved manuscript.The authors thank Richy Brown and Iain Neill for helpful discussions.
\end{acknowledgments}

\bibliographystyle{agufull08}
\bibliography{lun}

\end{article}
\end{document}